\documentclass[aps, prl, twocolumn, groupedaddress, reprint, floatfix ,longbibliography]{revtex4-2}
\usepackage{amsmath, amssymb, amsfonts}
\usepackage{bm}
\usepackage{graphicx}
\usepackage{hyperref}
\usepackage{braket}
\usepackage{color}
\usepackage{fontawesome}
\usepackage{orcidlink}

\definecolor{reviewpurple}{rgb}{0.5,0.0,0.5}
\definecolor{REVIEWPURPLE}{rgb}{0.5,0.0,0.5}

\newcommand{\github}[1]{\href{#1}{\faGithubSquare}}

\DeclareMathOperator*{\argmax}{argmax}

\begin{document}

\title{Are Molecules Magical?\\[0.2em]
{\small Non-Stabilizerness in Molecular Bonding}}

\author{Matthieu Sarkis\orcidlink{https://orcid.org/0009-0002-5494-8406}}
\email[]{matthieu.sarkis@uni.lu}
\affiliation{Department of Physics and Materials Science, University of Luxembourg, L-1511 Luxembourg City, Luxembourg}

\author{Alexandre Tkatchenko\orcidlink{https://orcid.org/0000-0002-1012-4854}}
\email[]{alexandre.tkatchenko@uni.lu}
\affiliation{Department of Physics and Materials Science, University of Luxembourg, L-1511 Luxembourg City, Luxembourg}

\date{\today}

\begin{abstract}
Isolated atoms as well as molecules at equilibrium are presumed to be simple from the point of view of quantum computational complexity.
Here we show that the process of chemical bond formation is accompanied by a marked increase in the quantum complexity of the electronic ground state. By studying the hydrogen dimer H$_{2}$ as a prototypical example, we demonstrate that when two hydrogen atoms form a bond, a specific measure of quantum complexity exhibits a pronounced peak that closely follows the behavior of the binding energy. This measure of quantum complexity, known as \textit{magic} in the quantum information literature, reflects how difficult it is to simulate the state using classical methods. We show that the observations for H$_{2}$ also hold for a collection of other dimers, including the weakly bonded diatomic helium dimer He$_{2}$. This observation suggests that regions of strong bonding formation or breaking are also regions of enhanced intrinsic quantum complexity. This insight suggests a connection of quantum information measures to chemical reactivity and advocates the use of stretched molecules as a quantum computational resource.
\end{abstract}

\maketitle

\section{Introduction}
    Quantum entanglement is a fundamental feature of many-body physics and quantum chemistry, reflecting nonclassical correlations between constituents \cite{rissler2006measuring,boguslawski2015orbital,ding2025entanglement}. It has become a key diagnostic in a wide range of systems—from condensed matter to molecular bonds—often quantified by measures such as the von Neumann entropy or Renyi entropies of reduced density matrices \cite{szalay2017correlation}. For example, in molecular systems such as the hydrogen molecule H$_2$, entanglement between electrons is negligible when the atoms are far apart or nearly fused into a single nucleus, but grows as a covalent bond forms. However, it is now understood that entanglement alone is not a sufficient indicator of a quantum state’s computational complexity or \textit{quantumness} \cite{bravyi2005universal,howard2014contextuality}. There exist highly entangled yet classically simulable states, notably the so-called \textit{stabilizer states} that can be produced by Clifford gates—transformations belonging to the normalizer of the Pauli group—and efficiently simulated on a classical computer via the Gottesman-Knill theorem \cite{gottesman1997stabilizer, gottesman1998theory, aaronson2004improved}. In other words, entanglement by itself does not guarantee quantum advantage. The resource that elevates a quantum state beyond stabilizer dynamics is referred to as \textit{non-stabilizerness} or \textit{magic}, as described by the resource theory of stabilizer quantum computation \cite{bravyi2005universal, bravyi2012magic, howard2014contextuality}. Stabilizer states and Clifford operations are considered \textit{free} since they can be simulated efficiently on a classical computer, whereas non-stabilizer states provide the essential resource for transcending classical simulability. Magic can be quantified by various monotones that do not increase under stabilizer operations; one convenient measure is the \textit{mana}, defined via the negativity of the state’s discrete Wigner function \cite{howard2014contextuality}. Indeed, stabilizer states possess a positive Wigner function, and for pure states, mana vanishes if and only if the state is a stabilizer state—a discrete analog of Hudson’s theorem linking positive Wigner representations to Gaussian states, thereby underlining that negativity in the quasiprobability distribution is essential for a state to supply quantum computational advantage. Besides mana, other magic measures have been proposed, such as the \textit{robustness of magic} \cite{howard2017application,heinrich2019robustness,hamaguchi2024handbook} and the \textit{stabilizer Renyi entropies} \cite{leone2022stabilizer, haug2023stabilizer}. Therefore, magic indicates the \textit{quantum overhead} inherent in accurately modeling a system. If a molecular system has low magic, then a classical method might suffice. High magic, on the other hand, flags the need for quantum computational resources.

    In parallel, quantum information concepts have begun to permeate the study of molecules and chemical bonding. The process of bond formation involves superposition of atomic configurations and entanglement between electrons. Measures such as entanglement entropy have been used to characterize these correlations in chemical systems. Rissler, Noack, and White \cite{rissler2006measuring} applied quantum information theory in chemistry by introducing orbital mutual information as a measure of electron interactions between orbitals, a concept that not only successfully identifies chemical bond patterns but also aids in optimizing DMRG algorithms. Building on this foundation, Boguslawski et al. \cite{boguslawski2015orbital} further developed the approach by calculating one- and two-electron entropies for molecular wavefunctions, thereby providing a more intuitive picture of electron correlation that informs the selection of active spaces. Recognizing the complexity of electron interactions, Ding et al. \cite{ding2020concept} refined these ideas by disentangling total orbital correlations into distinct classical and quantum components, which raised important questions regarding the genuine role of entanglement in chemical bonds. In a complementary effort, Szalay et al. \cite{szalay2017correlation} introduced a multiorbital correlation framework that utilizes genuine multipartite entanglement measures and clustering algorithms to reveal multi-center bonding patterns and to highlight the limitations of traditional bonding descriptions. Extending these insights to a more nuanced bonding analysis, Ding, Matito, and Schilling \cite{ding2025entanglement} proposed the concept of maximally entangled atomic orbitals (MEAOs), demonstrating that entanglement patterns can capture both conventional two-center bonds and delocalized multicenter bonds, with the degree of multipartite entanglement serving as a quantitative index of bond strength and aromaticity. Complementing these theoretical advances, Stein and Reiher \cite{stein2016automated} developed an automated protocol for active orbital space selection in multireference calculations, effectively leveraging entanglement measures to identify strongly correlated orbitals and streamline computational processes. Finally, hydrogen-based systems were analyzed~\cite{D4FD00066H} using diagnostics derived from the cumulant of the two-body reduced density matrix, showing that such quantities capture electronically nontrivial regimes along bond stretching and in correlated hydrogen aggregates.

    Let us insist though that the two resource theories of quantum correlations and non-stabilizerness, though not `orthogonal' in the space of resource theories, are not aligned. Our present work is complementary in spirit to the previously mentioned works in that it focuses instead on fermionic non-stabilizerness as a distinct quantum-information-theoretic notion.

    The goal of this work is therefore to combine these quantum information non-stabilizerness insights with quantum chemistry. We conduct an analysis of the non-stabilizerness in the H$_2$ molecule as it forms and breaks a bond. By doing so, we aim to illustrate how concepts like magic, alongside more standard notions like entanglement together provide a more complete characterization of the electronic wavefunction’s quantum nature. The rest of the paper is organized as follows. We first summarize the theoretical background and definitions of the various magic proxies used in this letter in the context fermionic systems. We then describe our methodology for computing our reference \textit{ab initio} ground state of the H$_2$ dimer accross a range of interatomic distances. We finally present the results, showing the behavior of magic as a function of interatomic distance, and provides a discussion of our observations, and discuss in some possible experimental implications in the outlook section.

\section{The Fermionic Wigner Function and Proxies of Magic}

    Non-stabilizerness concepts have be explored mostly in quantum spin chain context \cite{goto2021chaos, oliviero2022magic, smith2406non, tirrito2024anticoncentration, turkeshi2025magic, odavic2024stabilizer, passarelli2024nonstabilizerness, tarabunga2311magic, tarabunga2024critical}. These notion can of course be defined for fermionic systems in various defferent ways. One possible trajectory is fermionization in the form of Jordan-Wigner transformation from a system of spin 1/2 degrees of freedom, following then the work of Wootters \cite{wootters1987wigner, gibbons2004discrete}. Another direction could be to define a Grassmann valued phase space, and a notion of fermionic Wigner function defined thereof, following the seminal work of Cahill and Glauber \cite{cahill1999density}. We will instead follow the approach of \cite{mclauchlan2022fermion, mudassar2024encoding, collura2024quantum, bera2502non} to leverage the structure of the Majorana group \cite{bettaque2024structure}, fermionic analogue of the Pauli group, and define the fermionic Wigner function in terms of Majorana strings and discrete phase space.

    Given a collection of $n$ fermionic creation and annihilation operators $c_p$ and $c_p^\dagger$, following \cite{bettaque2024structure, bera2502non} we introduce for each mode the Hermitian Majorana operators $\eta_{2p-1} = c_p + c_p^\dagger$ and $\eta_{2p} = i(c_p - c_p^\dagger)$. We then define the Majorana strings
    \begin{equation}
        M_v = i^{v\cdot \Omega v} \eta_{1}^{v_{1}}\eta_{2}^{v_{2}}\dots\eta_{2n-1}^{v_{2n-1}}\eta_{2n}^{v_{2n}}\,,
    \end{equation}
    where $v = (v_1, v_2, \dots, v_{2n-1}, v_{2n})^\textsc{t} \in (\mathbb Z_2)^{2n}$ is a binary vector, and $\Omega$ is a $(2n)\times(2n)$ square matrix with zeros on the diagonal, zeros on the upper-right triangle, and ones on the lower-left triangle. The prefactor $i^{v\cdot \Omega v}$ ensures Hermiticity of the Majorana strings. $\Gamma=(\mathbb Z_2)^{2n}$ plays here the role of discrete phase space for the fermionic system. The Majorana strings form a basis of Hermitian operators, and given a quantum state represented by a density matrix $\rho$, one can decompose it in the Majorana basis as\footnote{This definition mimicks that of the Pauli spectrum in the case of qubit systems \cite{turkeshi2025pauli}.}
    \begin{equation}
        \rho = \frac{1}{2^{2n}}\sum_{v\in\Gamma}\text{Tr}(\rho M_v)M_v.
    \end{equation}
    We then call the quantity
    \begin{equation}
        W_\rho(v) = \text{Tr}(\rho M_v)
    \end{equation}
    the fermionic Wigner function of the state $\rho$.

    Given this fermionic Wigner function we define its $L^p$ norm as
    \begin{equation}
        || W_\rho ||_p = \left[\sum_{v\in\Gamma} | W_\rho(v) |^p\right]^{\frac{1}{p}}.
    \end{equation}
    The $\alpha$-stabilizer Renyi entropy is defined as
    \begin{equation}
        \mathcal S_\alpha = \frac{1}{1-\alpha}\log\left[\frac{|| W_\rho ||_{2\alpha}^{2\alpha}}{2^{2n}}\right].
    \end{equation}
    Following intuition from the discrete Wigner function of Wootters \cite{wootters1987wigner, gibbons2004discrete}, we define the mana as the $L^1$ norm instead:
    \begin{equation}
        \mathcal M = \log\left[\frac{|| W_\rho ||_{1}}{2^{2n}}\right].
    \end{equation}
    The filtered $\alpha$-stabilizer Renyi entropy $\mathcal{FS}_\alpha$ is defined like the $\alpha$-stabilizer Renyi entropy but removes the often dominating contribution of the identity $v = (0, 0, \dots, 0)^\textsc{t}$ and parity operator $v = (1, 1, \dots, 1)^\textsc{t}$ from the sum defining the $L^{p}$ norm \cite{collura2024quantum}, whose contribution can become dominant, especially in the large number of modes limit.

    The $L^p$ norms of the discrete Wigner function of a fermionic state capture how broadly the quantum state spreads over the discrete phase space $\Gamma$.

\section{The \texorpdfstring{H$_2$}{H2} dimer in second quantization: computing the ground state across dissociation}

    In the second quantization formalism, the electronic Hamiltonian (within the Born-Oppenheimer approximation) of a molecular system is expressed in terms of fermionic creation (\(c_p^\dagger\)) and annihilation (\(c_q\)) operators defined with respect to a chosen orbital basis. For a generic set of orbitals, the Hamiltonian is written as~\footnote{The constant nuclear repulsion energy is omitted for simplicity, but of course contributes to the total energy of the system.}
    \begin{equation}
    \hat{H} = \sum_{p,q} h_{pq}\, c_p^\dagger c_q + \frac{1}{2} \sum_{p,q,r,s} \langle pq \vert rs \rangle\, c_p^\dagger c_q^\dagger c_s c_r,
    \end{equation}
    where $h_{pq}$ are the one-body integrals (incorporating the kinetic energy of electrons and their interaction with the nuclei), $\langle pq\vert rs\rangle$ are the two-body integrals (accounting for electron-electron repulsion), and the quantum numbers $p, q, r, s$ run over the complete set of orbitals in the basis \cite{mcweeny1989method, cramer2013essentials}. This non-relativistic field theory representation encapsulates all the many-body effects and provides a convenient framework for \textit{ab initio} quantum chemical calculations\footnote{For non-quantum chemists, we recommend the very nice paper \cite{graves2023electronic} for a compact and self-contained exposition of some standards \textit{ab initio} methods.}  \cite{szabo1996modern, helgaker2013molecular}.

    For the hydrogen dimer H$_2$, we adopt the minimal STO-3G basis set, where each hydrogen atom is described by a linear combination of three Gaussian functions approximating the 1s atomic orbital. Despite its simplicity, the STO-3G basis offers a tractable, yet accurate model for exploring fundamental electronic properties of the hydrogen dimer.

    To accurately determine the ground state of H$_2$, we adopt a Full Configuration Interaction (FCI) approach on top of a Hartree--Fock mean-field solution, allowing different spatial orbitals for electrons of different spins ($\alpha$ and $\beta$), which is crucial for spin contamination \cite{szabo1996modern, helgaker2013molecular} considerations. This is particularly important in the dissociation limit in order to describe the correct covalent bond breaking and yield physically accurate results. We solve for the electronic Schrödinger equation exactly within the STO-3G basis set \cite{hehre1969self}. FCI therefore provides our benchmark for electron correlation, ensuring an accurate description of the ground state of the system across all interatomic distances.

\section{Results and Discussion}

\begin{figure}[ht]
        \centering
            \includegraphics[width=0.5
            \textwidth]{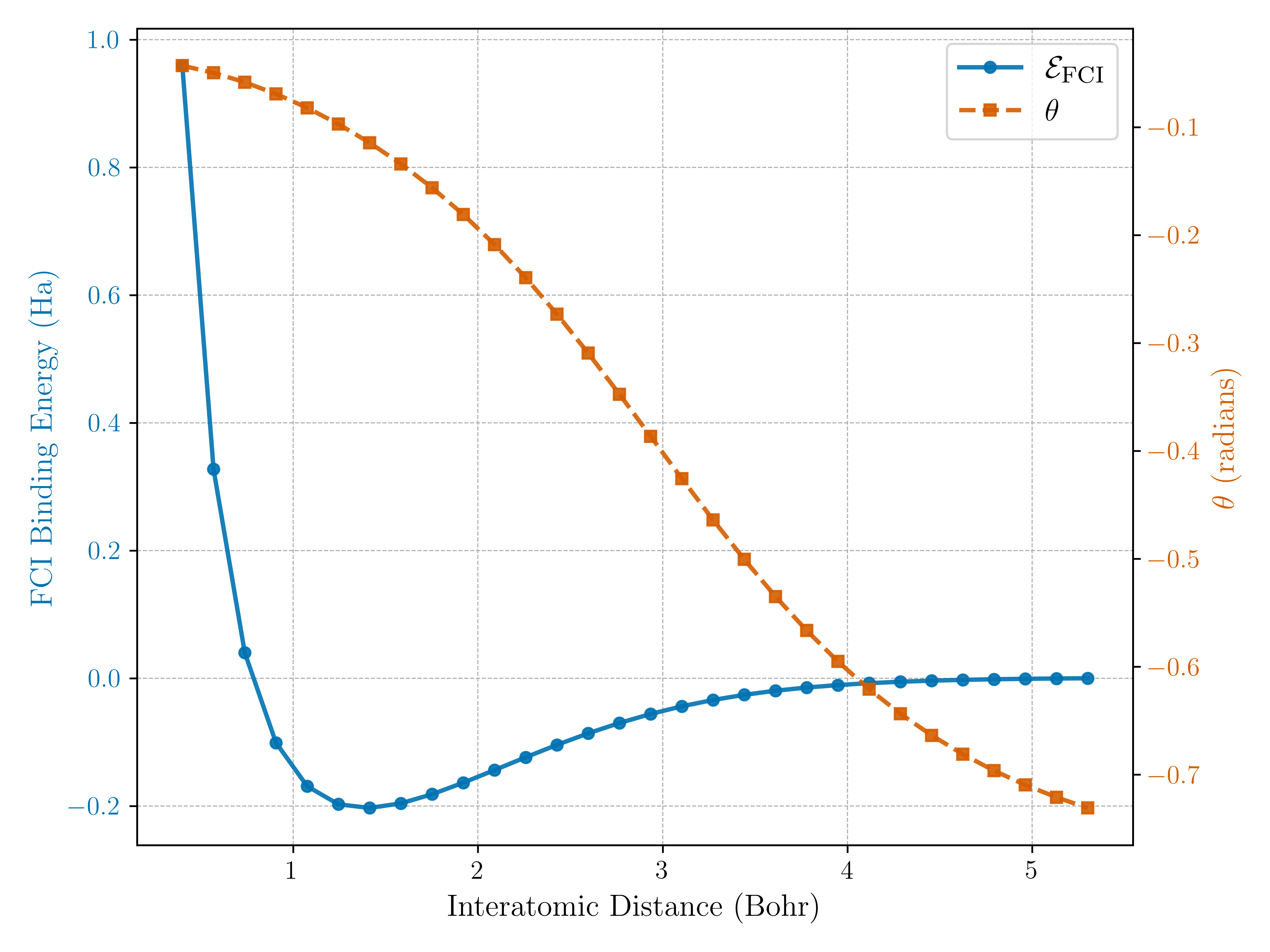}
            \caption{FCI binding energy $\mathcal E_\textsc{fci}$ and $\theta$ angle defining the ground state wavefunction of the H$_2$ dimer as a function of the interatomic distance.}
    \label{fig:binding_theta}
    \end{figure}

    The FCI ground state of the system is very well captured at any interatomic distance by a state of the form
    \begin{equation}
    \label{eq:ground_state}
        |\psi(\theta)\rangle = \cos(\theta)\,|1100\rangle + \sin(\theta)\,|0011\rangle
    \end{equation}
    with the ordering of the four fermionic modes given by: (i) spin $\alpha$ molecular orbital 0, (ii) spin $\beta$ molecular orbital 0, (iii) spin $\alpha$ molecular orbital 1, and (iv) spin $\beta$ molecular orbital 1. The second determinant corresponds to the fully excited state whose contribution is crucial for avoiding spin contamination in the large interatomic distance limit. The angle $\theta\equiv\theta(\ell)$ is a smooth function of the interatomic distance $\ell$, connecting the large distance limit in which the system factorizes into a pair a independent hydrogen atoms to the interatomic distance regime of covalent bond formation. Indeed, at these large distances, the purely ionic contribution from the two determinants precicely cancel each other, leaving solely the purely covalent contribution. Around the bound state, the Hartree-Fock contribution alone provides instead a qualitatively good description of the ground state. The reader will find in Fig.~\ref{fig:binding_theta} the binding energy curve of the H$_2$ molecule, namely the FCI energy
    \begin{equation}
        \mathcal E_\textsc{fci}(\ell) = \left\langle\psi(\theta(\ell))\left|\hat{H}\right|\psi(\theta(\ell))\right\rangle
    \end{equation}
    as a function of the interatomic distance, translated by the asymptotic contribution of two isolated hydrogen atoms $\mathcal E_\textsc{fci}(\infty)$, as well as the angle $\theta$ defining the corresponding FCI ground state wavefunction.

    \begin{figure*}[ht]
    \centering
        \includegraphics[width=0.85
        \textwidth]{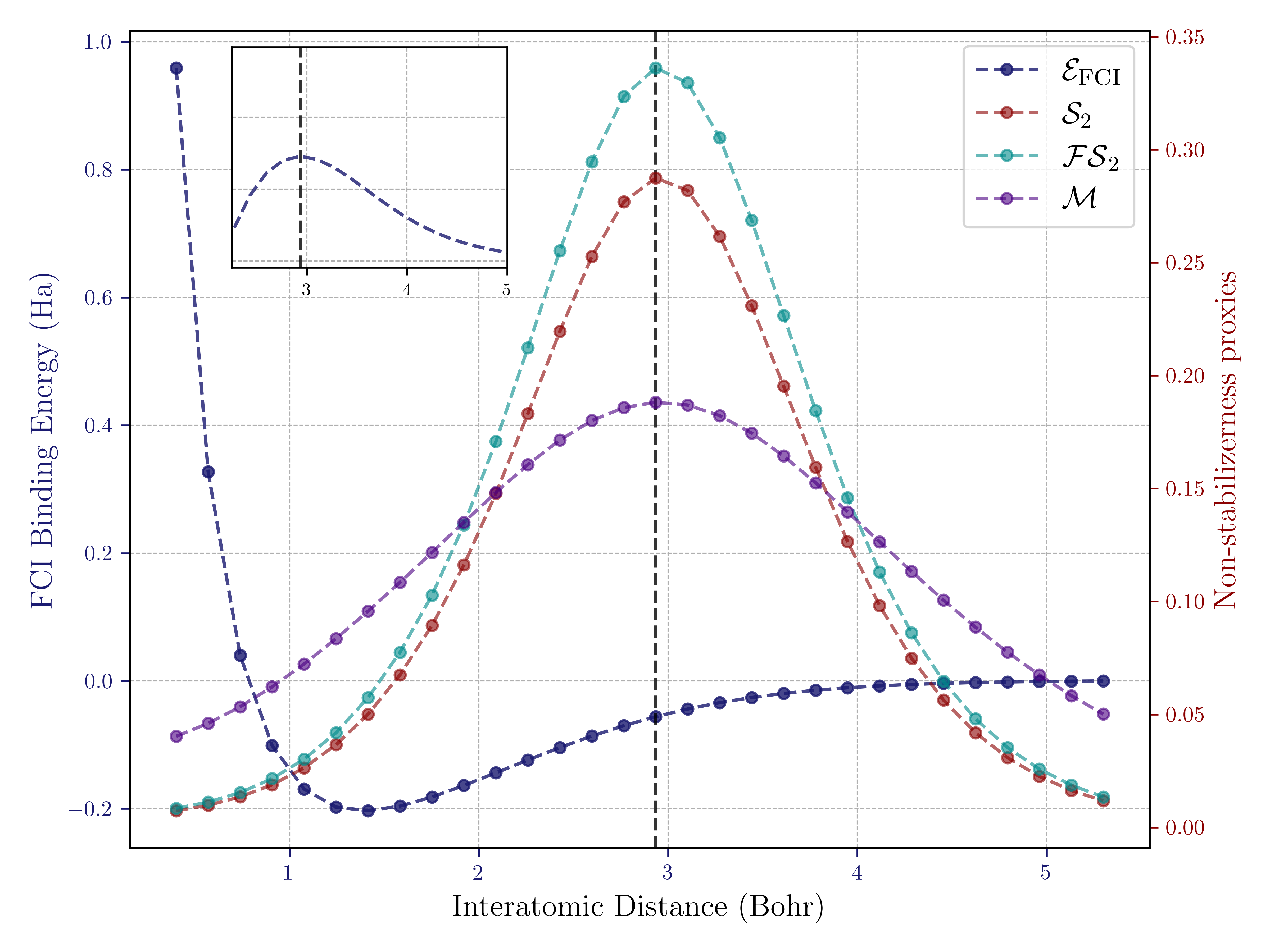}
        \caption{Stabilizer Renyi entropy $\mathcal S_2$, filtered stabilizer Renyi entropy $\mathcal{FS}_2$, mana $\mathcal M$, and FCI binding energy $\mathcal E_\textsc{fci}$ as a function of the interatomic distance. These magic proxies exhibit a pronounced peak precicely at the value of the interatomic distance where the extrinsic curvature of the binding energy curve is extremized, as indicated by the vertical dashed line. The inset now directly depicts the extrinsic curvature $\kappa(\ell)$ of the FCI binding energy curve, making the correspondence with the marked reference distance explicit.}
    \label{fig:magic_vs_distance}
    \end{figure*}

\begingroup
\refstepcounter{table}
\noindent\textbf{Table \thetable.} Numerical values corresponding to Fig.~\ref{fig:magic_vs_distance}. The binding energy is $\mathcal E_\textsc{fci}(\ell^\star)-\mathcal E_\textsc{fci}(\infty)$.\label{tab:magic_vs_distance}
\begin{center}
\begin{tabular}{lc}
\hline
Basis & STO-3G \\
$\ell^\star$ (Bohr) & 2.934 \\
$\mathcal E_\textsc{fci}(\ell^\star)-\mathcal E_\textsc{fci}(\infty)$ (Ha) & -0.05592 \\
$\mathcal S_2(\ell^\star)$ & 0.28748 \\
$\mathcal{FS}_2(\ell^\star)$ & 0.33623 \\
$\mathcal M(\ell^\star)$ & 0.18818 \\
\hline
\end{tabular}
\end{center}
\endgroup

    Let us define the extrinsic curvature of the binding energy curve as
    \begin{equation}
        \kappa(\ell) = \frac{\left|\mathcal E_\textsc{fci}''(\ell)\right|}{\left(1+\mathcal E_\textsc{fci}'(\ell)^2\right)^{3/2}},
    \end{equation}
    Let us denote by $\ell^\star$ the point of maximal extrinsic curvature of the binding energy curve $\ell^\star = \argmax_{\ell} \kappa(\ell)$.
    For comparison with a more traditional mean-field marker, one may also define the Coulson--Fischer point as the first interatomic distance at which the unrestricted Hartree--Fock solution becomes lower in energy than the restricted Hartree--Fock one, signaling the onset of spin-symmetry breaking at the mean-field level. In our STO-3G setup, this yields $\ell_\textsc{cf}\simeq 2.14$ Bohr, which lies noticeably to the left of our exact curvature point $\ell^\star \simeq 2.93$ Bohr \footnote{Interestingly, the same STO-3G scan places the inflection point of the exact FCI binding curve at $\ell_\mathrm{infl}\simeq 2.13$ Bohr, i.e.\ almost on top of $\ell_\textsc{cf}$. We do not investigate the near-coincidence between $\ell_\textsc{cf}$ and $\ell_\mathrm{infl}$ further here, since that would take us too far from the main topic of the paper.}. This estimate is also compatible with the classic H$_2$ Coulson--Fischer analysis, which placed the breakdown of the symmetric molecular-orbital picture near $R\simeq 2.27$ Bohr, and with modern unrestricted-Hartree--Fock discussions of the same symmetry-breaking regime in stretched H$_2$ \cite{coulson1949hf,hait2019cf}. The two quantities therefore capture related but distinct physical phenomena: the former marks the onset of mean-field symmetry breaking, whereas the latter identifies the point of maximal extrinsic curvature of the exact FCI binding curve.
    For each value of the interatomic distance $\ell$, we compute the fermionic Wigner function of the ground state and evaluate the magic proxies defined in the first Section of the paper. The reader will find in Fig.~\ref{fig:magic_vs_distance} the behavior of the stabilizer Renyi entropy $\mathcal S_2$ and the mana $\mathcal M$ as a function of the interatomic distance $\ell$. The corresponding numerical values at $\ell^\star$ are collected in Table~\ref{tab:magic_vs_distance}.

    Our results reveal a striking phenomenon: as the hydrogen atoms approach each other, the magic proxies develop a pronounced peak in the same intermediate stretching regime in which the extrinsic curvature of the binding energy curve is maximal \footnote{In fact, within the exact STO-3G minimal-basis model we can rigorously prove that the two locations differ by only $\ell_{\mathrm{magic}}^{\mathrm{peak}}-\ell^\star \simeq 0.0423$ Bohr, i.e.\ about $0.0224$ \AA. They are therefore distinct, but chemically very close.}. This suggests that the bonding process is accompanied by a significant increase in non-stabilizerness, implying that the formation of a covalent bond requires the consumption of a large amount of non-Clifford operations. A distinct short-distance curvature maximum also appears in the strongly repulsive regime, but it is not accompanied by a comparable enhancement of the magic proxies. In that compressed regime, the ground state remains close to a weak-magic closed-shell configuration, so the large curvature mainly reflects the steep energetic cost of nuclear repulsion rather than a rise in non-stabilizerness. As we discussed, a state with low magic is efficiently classically simulable, while high magic is a necessary ingredient for quantum computational speedup \cite{howard2014contextuality,Veitch2014Resource}. Thus, our analysis indicates that the H$_2$ bond formation is not only a chemical process but also a transformation that incurs a cost in terms of quantum computational resources.

    The angle $\theta$ defining the ground state wavefunction adjusts smoothly with $\ell$, reflecting the relative weight of two determinants. Equal determinant weights correspond instead to the dissociation limit, where $\theta\to -\pi/4$. The high-magic point occurs earlier along the dissociation curve, near $\theta=-\pi/8$, where both determinants contribute non-equally but both non-negligibly. In that intermediate regime, the interference between them amplifies off‑diagonal correlations in the Majorana basis, leading to a more \textit{spread out} fermionic Wigner function and, consequently, to a higher value of our magic proxies as defined by $L^p$ norms of the fermionic Wigner function.

    From the standpoint of molecular physics and quantum chemistry, our results offer an intriguing reinterpretation of chemical bond formation: in the dissociation limit ($\ell\to\infty$), the hydrogen atoms are essentially isolated. In this limit, the Hamiltonian separates into two non-interacting atomic contributions, and the electronic state reduces to the purely covalent dissociation-limit. In such a regime, the fermionic correlations are minimal and the associated magic (or non‑stabilizerness) is low, consistent with the notion of a classically tractable system.

    At interatomic distances near $\ell^\star$, where the binding energy curve shows maximal extrinsic curvature, the wavefunction represents a delicate balance between the two determinants. This is the regime where the chemical bond is forming. The competition between the ionic and covalent contributions results in a highly correlated state. Our analysis shows that this neighborhood is also the one in which the quantum state demands the largest number of non‑Clifford resources for its simulation, as reflected by the nearby peak in the magic proxies. The extrinsic curvature $\kappa$ is a geometric measure of the sensitivity of the binding energy with respect to interatomic distance. Its maximum marks a rapid change in the energy landscape, signature of a transition in the electronic structure. This energetic reorganization is directly correlated with the rise in non‑stabilizerness, indicating that the very formation of the covalent bond is accompanied by an increase in the \textit{quantumness} of the state.

    Very interestingly, note that the ground state (\ref{eq:ground_state}) can be understood as the state of a qubit upon interpreting the two determinants as computational basis states $\left|\tilde 0\right\rangle=|1100\rangle$ and $\left|\tilde 1\right\rangle=|0011\rangle$. We know that at very short interatomic distances, including the equilibrium bound state distance, the ground state of the system is very well approximated by setting $\theta\simeq 0$, namely by the state $\left|\tilde 0\right\rangle$. Adiabatic increase of the bond length can then be interpreted as implementing the following unitary operation:
    \begin{equation}
        |\psi(\theta)\rangle = U(\theta)\left|\tilde 0\right\rangle = \cos(\theta)\left|\tilde 0\right\rangle + \sin(\theta)\left|\tilde 1\right\rangle\,,
    \end{equation}
    with $U(\theta)$ the rotation around the $y$-axis
    \begin{equation}
        U(\theta) = R_y(2\theta) = \begin{pmatrix}
            \cos\theta & -\sin\theta \\ \sin\theta & \cos\theta
        \end{pmatrix}\,.
    \end{equation}
    Note that one can conjugate this rotation into a rotation around the $z$-axis instead:
    \begin{equation}
        U(\theta) = R_x\left(\frac{\pi}{2}\right)R_z(2\theta)R_x\left(-\frac{\pi}{2}\right)\,.
    \end{equation}
    As explained below, the analytic maximum of the minimal-basis magic proxies occurs at $\theta=-\pi/8$, which corresponds to an interatomic distance very close to the curvature reference point $\ell^\star$ in our STO-3G scan, though not exactly equal to it. The key observation is the following: for $\theta=-\pi/8$, the unitary $U(-\pi/8)$ is precisely conjugate to the T-gate (more precisely to its Hermitian conjugate in our conventions), and \textit{the conjugation matrix belongs to the Pauli group}. The T-gate is precisely known to be a non-Clifford gate which once adjoined to the group of Clifford operations allows for universal quantum computation. Morevover, conjugation by an element of the Pauli group does not impact non-stabilizerness, and therefore adiabatically stretching the bond from equilibrium into this high-magic intermediate regime can be viewed, at least qualitatively, as implementing a highly non-Clifford transformation. Conversely, relaxing from that regime back toward equilibrium may be interpreted in the same spirit as undoing such a transformation. We push this reasoning further at the end of the outlook section.

    The confluence of these perspectives reinforces a remarkable insight: the process of chemical bond formation is not solely an energetic or structural rearrangement but is also accompanied by a non‑trivial transformation in the quantum informational character of the state. At large distances, the electrons are described by nearly independent, stabilizer‑like states, whereas in the bonding formation region the superposition of covalent and ionic contributions requires an injection of magic into the system. This observation opens a conceptual bridge between quantum resource theories and chemical reactivity, suggesting that the cost of forming a bond can be viewed through the lens of quantum computational resources.

    \section{Entanglement Entropy and Bonding Correlations}

\begin{figure}[ht]
        \centering
        \includegraphics[width=0.5\textwidth]{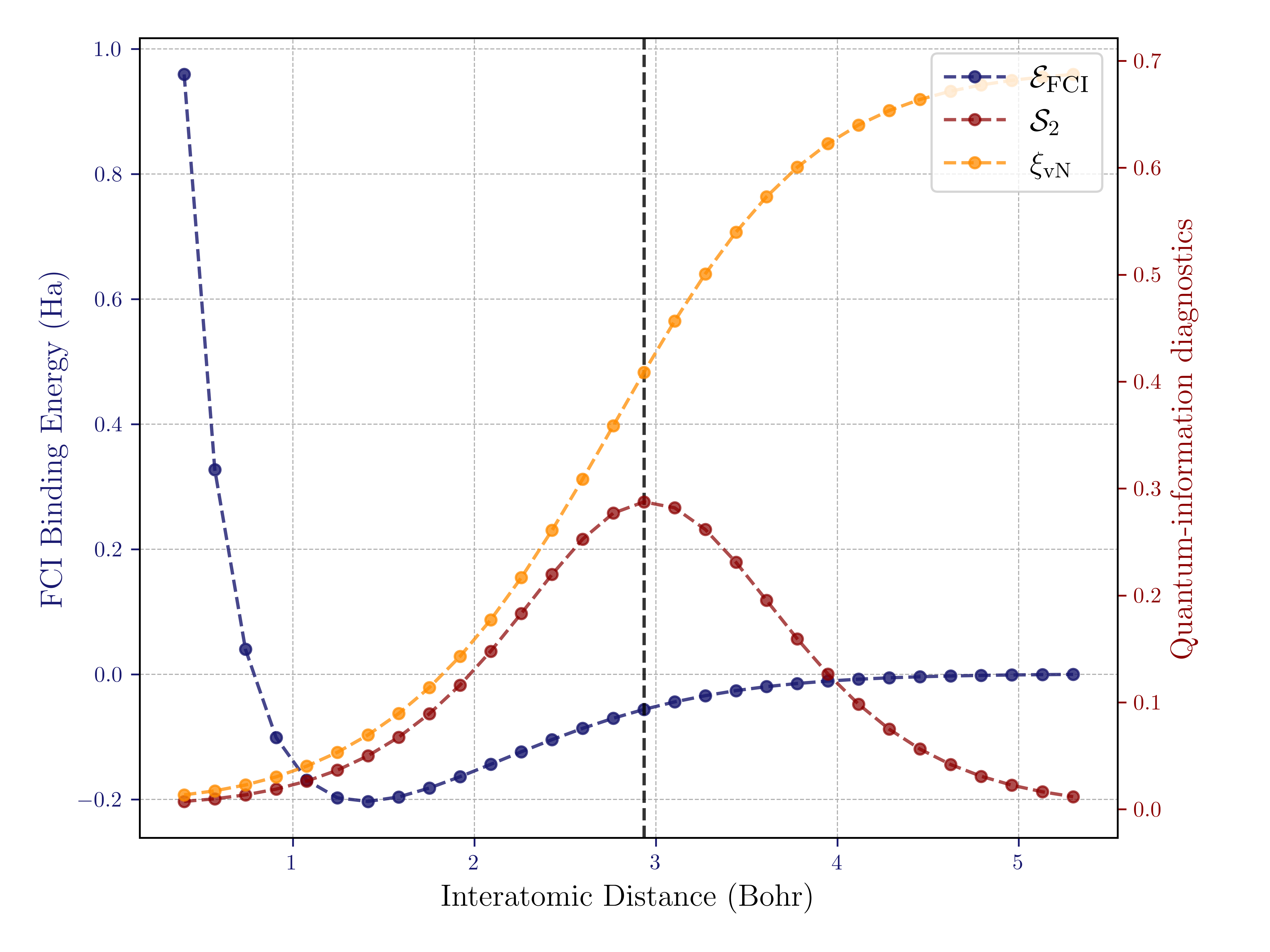}
        \caption{FCI binding energy $\mathcal E_\textsc{fci}$, stabilizer R\'enyi entropy $\mathcal S_2$, and electronic entanglement entropy $\xi_{\mathrm{vN}}$ of Ref.~\cite{esquivel2011quantum}, as functions of the interatomic distance in the STO-3G basis. The vertical dashed line marks the bond-formation point $\ell^\star$ used throughout the manuscript.}
        \label{fig:magic_entanglement_vs_distance}
    \end{figure}

    To compare our magic diagnostics with a more standard correlation measure, we also evaluated the \emph{electronic} entanglement measure used by Esquivel \textit{et al.}\ \cite{esquivel2011quantum}. It is defined from the one-particle reduced density matrix $\rho_{\textsc{rdm}}(\ell)$ of the correlated electronic state:
    \begin{equation}
        \xi_{\mathrm{vN}}(\ell)=-\mathrm{Tr}\left[\rho_{\textsc{rdm}}(\ell)\log \rho_{\textsc{rdm}}(\ell)\right],
    \end{equation}
    which quantifies entanglement between the \emph{electrons}. The corresponding behavior is shown in Fig.~\ref{fig:magic_entanglement_vs_distance}, where we plot the FCI binding energy together with the stabilizer R\'enyi entropy $\mathcal S_2$ and the electronic entropy $\xi_{\mathrm{vN}}$.

    The comparison reveals a clear qualitative difference between these diagnostics. The electronic entropy $\xi_{\mathrm{vN}}$ is small near the united-atom limit and grows as the molecule enters the stretched, strongly correlated regime, approaching the dissociation value discussed in Ref.~\cite{esquivel2011quantum}. By contrast, our magic proxy $\mathcal S_2$ develops a pronounced maximum at an intermediate geometry and then decreases again as one moves toward dissociation.

    This distinction is important for the interpretation of our results. The electronic entanglement curve does not develop the sharp intermediate-distance maximum exhibited by the magic proxy $\mathcal S_2$: at the high-magic point near $\ell^\star$, $\xi_{\mathrm{vN}}$ is certainly nonzero, but it is not extremal there. The magic peak is therefore not a disguised version of ordinary entanglement growth. Rather, it singles out a more specific regime in which the correlated molecular wavefunction becomes maximally \emph{non-stabilizer}, i.e.\ maximally far from the efficiently classically simulable fermionic structures captured by stabilizer theory. A state may be strongly entangled while still having modest magic, and conversely the point of maximal magic need not coincide with the points emphasized by standard entanglement diagnostics. Our purpose here is not to undertake a systematic study of the various correlation-based approaches to bonding, but rather to underline the novelty of the information carried by magic measures with respect to more standard correlation measures. The novelty of the present work is precisely to isolate this additional layer of quantum structure in molecular bonding, beyond standard correlation measures such as entanglement entropy \cite{esquivel2011quantum}.

\section{Extension to Additional Dimers}

To test whether the qualitative picture identified for H$_2$ extends beyond that minimal case, we broaden the analysis to four additional dimers spanning distinct interaction regimes: LiH, HF, BeH$^+$, and He$_2$. For LiH, HF, and BeH$^+$, we use compact valence CASSCF(2,2) wavefunctions in the 6-31g basis and evaluate both the binding curve and the filtered stabilizer R\'enyi entropy $\mathcal{FS}_2$. Since the active spaces remain small, the filtered Majorana sum can again be computed straightforwardly.

\begin{figure*}[ht]
    \centering
    \includegraphics[width=\textwidth]{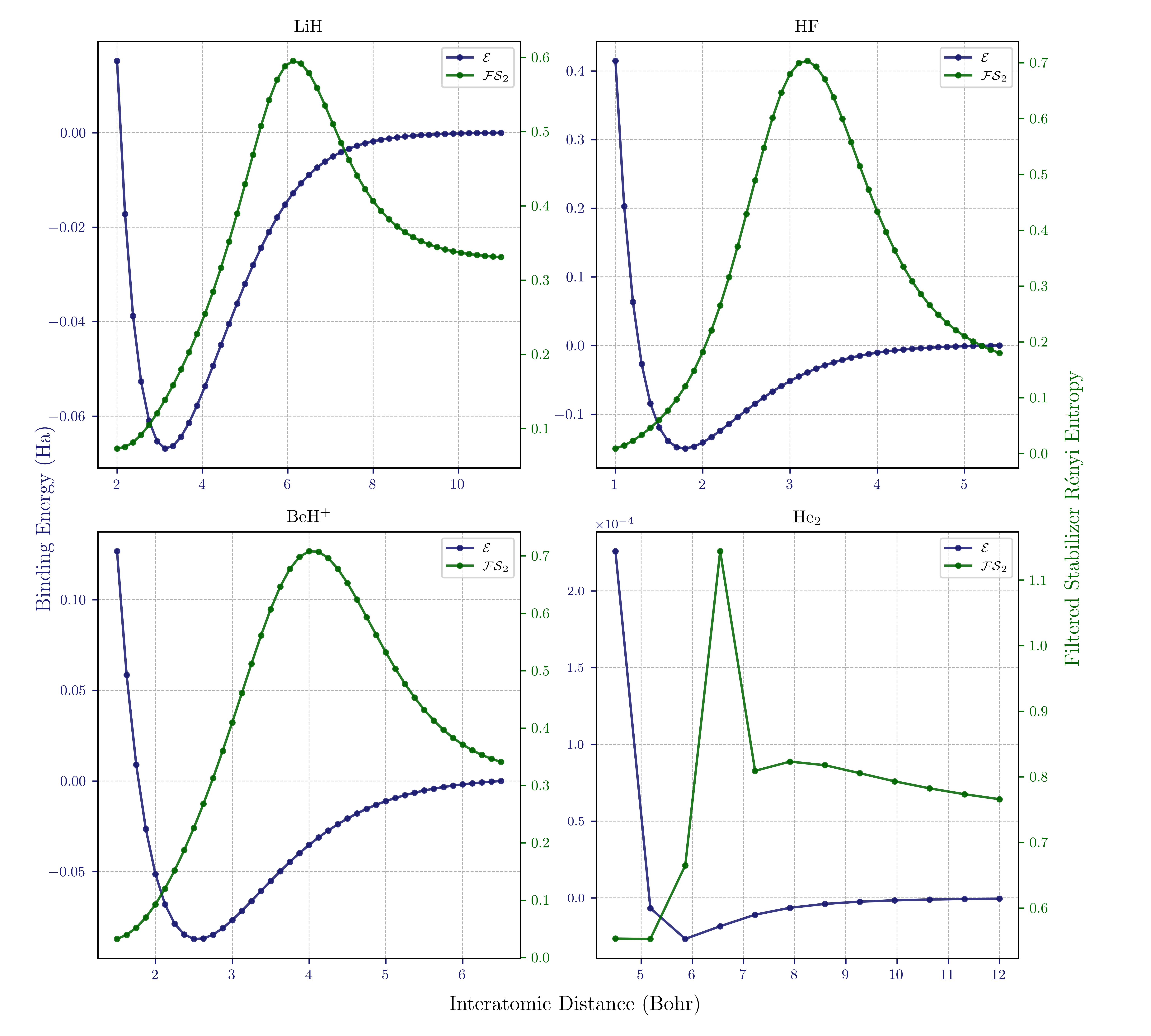}
    \caption{Binding energy and filtered stabilizer R\'enyi entropy $\mathcal{FS}_2$ for four additional dimers: LiH, HF, BeH$^+$, and He$_2$. For LiH, HF, and BeH$^+$, both curves are obtained from the same CASSCF(2,2) wavefunction in the 6-31g basis. The He$_2$ panel instead combines a high-level electronic-structure treatment of the very shallow van der Waals well with a correlated compact-wavefunction evaluation of $\mathcal{FS}_2$, and is therefore intended as a qualitative comparison across bonding scenarios rather than as a chemically precise benchmark.}
    \label{fig:magic_vs_distance_four_dimers}
\end{figure*}

The purpose of this comparison is not to provide chemically accurate benchmark descriptions of all these systems on equal footing. Rather, it is to test the \emph{generality} of the trend observed in H$_2$: namely, whether an enhancement of the magic proxy at intermediate separations also appears in chemically distinct two-body systems. This is also the reason for including He$_2$. As a weakly bound van der Waals dimer, it probes a qualitatively different interaction regime and therefore serves as a useful control case for scope.

As shown in Fig.~\ref{fig:magic_vs_distance_four_dimers}, all four systems exhibit a maximum of the magic proxy at an intermediate interatomic distance, although the detailed peak positions, energy scales, and asymptotic behaviors depend on the system. LiH, HF, and BeH$^+$ therefore suggest that the bond-stretching enhancement of non-stabilizerness is not peculiar to H$_2$, but persists across chemically distinct bonding situations. In LiH, for example, $\mathcal{FS}_2$ does not return to zero over the plotted range; within the present CASSCF(2,2)/6-31g description, the stretched-state wavefunction remains non-stabilizer in the chosen active-orbital basis. We do not regard that nonzero asymptote as the central point, since it may depend on the orbital and basis representation; the more robust feature is the presence of the intermediate-distance peak itself.

He$_2$ should be interpreted more cautiously. Because it is an extremely weakly bound van der Waals dimer, the compact active-space treatment used here is not intended to yield a quantitatively reliable description of its binding energetics. Chemical accuracy is therefore not the goal of this analysis. Rather, the point is to test whether the same magic phenomenology remains visible in a qualitatively different bonding regime. At that level, the answer appears to be affirmative. We therefore view these results as support for the broader generalizability of the mechanism observed in H$_2$.

\section{Remark concerning generalizability of the results to realistic basis sets, and analytical expression of the magic proxies}

\begin{figure}[ht]
    \centering
        \includegraphics[width=0.5
        \textwidth]{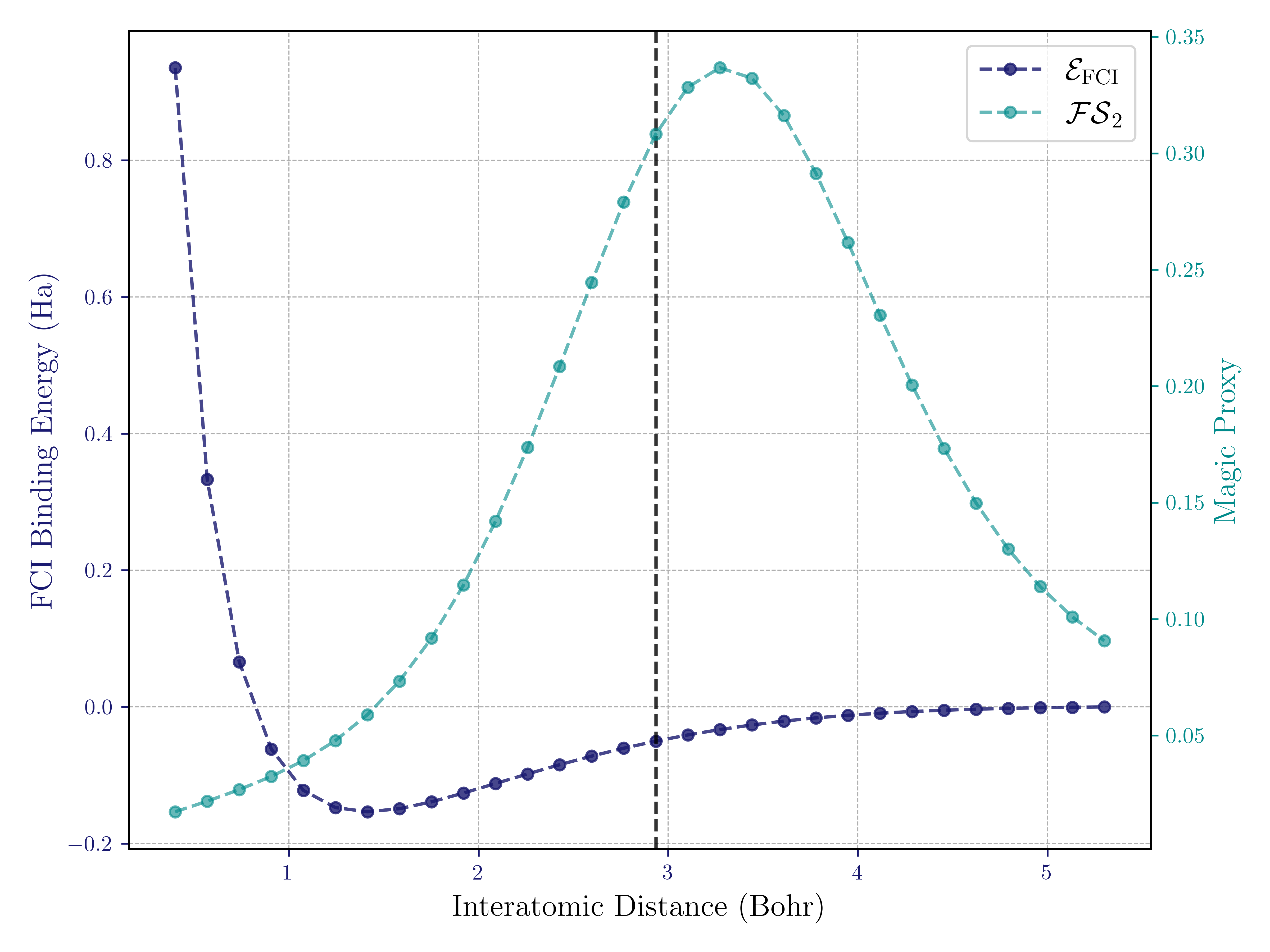}
        \caption{FCI binding energy $\mathcal E_\textsc{fci}$ and filtered stabilizer Renyi entropy in the 6-31g basis. The vertical dashed line marks again the maximal extrinsic curvature point $\ell^\star$. One observes that the maximum of $\mathcal{FS}_2$ is shifted to the right with respect to this reference point.}
    \label{fig:magic_vs_distance_631g}
    \end{figure}

    One can of course question whether the results observed above in the case of a minimal basis set generalize to richer, more expressive and realistic basis sets. Without entering into the technical details, let us show the results we obtain still in the case of the hydrogen dimer, but when employing the larger 6-31g basis set. The reader will find in Fig.~\ref{fig:magic_vs_distance_631g} the depiction of the FCI binding energy curve and of the filtered 2-stabilizer Renyi entropy.

    \begingroup
    \refstepcounter{table}
    \noindent\textbf{Table \thetable.} Numerical values corresponding to Fig.~\ref{fig:magic_vs_distance_631g}. These values are obtained from the same generated 6-31g data used to plot Fig.~\ref{fig:magic_vs_distance_631g}.\label{tab:magic_vs_distance_631g}
    \begin{center}
    \begin{tabular}{lc}
    \hline
    Basis & 6-31g \\
    $\ell^\star$ (Bohr) & 2.934 \\
    $\mathcal{FS}_2(\ell^\star)$ & 0.30820 \\
    $\ell^{\mathrm{peak}}_{\mathcal{FS}_2}$ (Bohr) & 3.272 \\
    $\mathcal{FS}_2^{\mathrm{peak}}$ & 0.33673 \\
    $\Delta \ell=\ell^{\mathrm{peak}}_{\mathcal{FS}_2}-\ell^\star$ (Bohr) & 0.338 \\
    \hline
    \end{tabular}
    \end{center}
    \endgroup

    We still observe a neat peak in the magic proxy. The maximum of the 6-31g filtered stabilizer Renyi entropy is found at a slightly larger interatomic distance, as shown in Fig.~\ref{fig:magic_vs_distance_631g} and Table~\ref{tab:magic_vs_distance_631g}. In the case of the hydrogen dimer, the use of a minimal basis set can already be considered relatively accurate, and as we saw it leads to a very close, though not exact, alignment between the point of maximal sensitivity, as characterized by the curvature data of the potential energy surface, and the point of maximal non-stabilizerness. This points towards a deep connection between the non-classicality of molecular systems and the Riemannian geometry of their potential energy surface. The observed slight correction brought to this statement by the use of a larger basis set can be understood in a first approximation by the following argument: the larger the available orbitals, the earlier the start of their overlap as the two atoms get closer and closer to each other.

    Let us also mention that in the case of a minimal basis set for H$_2$, the curse of dimensionality that plagues quantum chemistry for larger systems is absent, allowing for an analytic expression if the fermionic Wigner function, and therefore of the magic proxies. One can indeed easily see that among the 32 possible Majorana strings, 16 of them act diagonally in the space spanned by the two determinants, and 16 are off-diagonal and contribute to the cross term. Among the diagonal strings, depending on the relative contribution of the two determinants, 8 of them contribute $
    \cos^2(\theta)+\sin^2(\theta)=1$ and 8 of them contribute $
    \cos^2(\theta)-\sin^2(\theta)=\cos(2\theta)$. Concerning the off-diagonal strings, 8 of them give a zero contribution, and the other 8 contribute $2\sin(\theta)\cos(\theta)=\sin(2\theta)$. One can then extract the various magic proxies. For the stabilizer 2-Renyi entropy and its filtered counterpart one obtains:
    \begin{equation}
        \begin{aligned}
        \mathcal S_2(\theta) &= -\log\left[1 - \frac{\sin^2(4\theta)}{4}\right]\,, \\
        \mathcal{FS}_2(\theta) &= -\log\left[1 - \frac{2}{7}\sin^2(4\theta)\right]\,,
        \end{aligned}
    \end{equation}
    and for the mana:
    \begin{equation}
        \mathcal M(\theta) = \log\left[\frac{1+|\cos(2\theta)|+|\sin(2\theta)|}{2}\right]\,.
    \end{equation}
    We indeed check that the analytic maxima of these two quantities occur at $\theta=-\pi/8$, where they take the respective values $\log(4/3)\simeq 0.287(6)$ and $\log\left(\tfrac{1}{2}+\tfrac{1}{\sqrt 2}\right)\simeq 0.188(2)$, as can be directly checked in Fig.~\ref{fig:magic_vs_distance}. Though not scalable to more realistic basis sets and/or larger systems, this analytic result helps explain why the STO-3G magic peak lies very close to the maximal extrinsic-curvature point of the FCI binding curve, without requiring that the two coincide exactly.

\section{Outlook}

    While our study focuses on the simple hydrogen dimer, it would be instructive to extend this analysis to more complex molecules. The methodology we employed – combining \textit{ab initio} methods with quantum resource-theoretic measures – can be extended to other molecules and more sophisticated basis sets. One could, for instance, analyze the magic content in a stretched water molecule or in a transition metal dimer where multi-reference character is strong. We expect that systems requiring multi-reference descriptions will generally show non-zero mana.

    Furthermore, the minimal STO-3G basis, despite its simplicity, reproduces H$_2$ well depth with remarkable accuracy \cite{szabo1996modern, hehre1969self}, suggesting that essential bond-formation physics—balancing ionic and covalent contributions and rising quantum correlations—is already captured. Thus, the pronounced peak in magic proxies at bond formation should persist for heavier atoms and larger basis sets. Although such extensions may adjust quantitative details of the binding curve and magic measures, the qualitative link between non-stabilizerness and covalent bonding remains. A key obstacle to scaling is the exponential growth of Majorana strings (fermionic phase space), which can be mitigated via Monte Carlo estimates of magic proxies \cite{bera2502non}.

    The additional dimer examples suggest that the intermediate-distance peak of magic is not specific to H$_2$, but may persist across rather different bonding regimes, including weakly bound van der Waals systems. This motivates extending the analysis from selected dimers to broader molecular families, in order to determine which aspects of the effect are robust and which depend on the bonding mechanism. In particular, it would be valuable to study whether similar magic-enhancement patterns appear in larger polyatomics, along reaction coordinates, or near avoided crossings and strongly multireference regions, where electronic reorganization is significant.

    The observed interplay between electronic structure and quantum computational resources suggests that quantum simulations of chemistry could benefit from incorporating magic measures as diagnostic tools. Understanding how magic is generated and consumed during chemical reactions could indeed inform the design of more efficient quantum algorithms for molecular modeling. For instance these insights may lead to the design of more efficient quantum algorithms that adapt to the changing resource demands along reaction coordinates. Similar questions were discussed regarding fermionic systems mapped to qubit systems for simulations on near-term devices \cite{gu2024zero}.

    Although magic is a computational resource, indirect experimental signatures—such as shifts in entanglement spectra or spectroscopic features—may correlate with high magic. Exploring these could offer a new window into quantum correlations and non-stabilizerness in chemistry. Our results hint at a broader framework where quantum-information concepts beyond entanglement entropy (like non-stabilizerness) signal physical phenomena such as bonding. It would be illuminating to see how mana scales with system size—does adding electrons in similar bonds increase mana proportionally, or do mean-field effects dominate? Do magic-proxy peaks universally mark covalent bond formation or reaction barriers? Studying larger systems could reveal whether non-stabilizerness correlates with reactivity or catalytic efficiency.

    Finally, an intriguing perspective emerges when we consider harnessing the observed surge in non‑stabilizerness as a resource for quantum computation. Imagine a protocol in which a dimer, initially at its equilibrium position, is adiabatically stretched, perhaps via coupling to an external field, to deliberately enhance the non‑stabilizerness stored in its ground state. In this scenario, the dimer effectively serves as a reservoir of quantum magic that could later be coupled to an external device. Extending this idea to an array of dimers, one could control not only the spatial orientation and individual bond lengths but also the inter-dimer couplings, thereby engineering a system in which both non‑stabilizerness and quantum correlations are actively leveraged for computational tasks. Note that our observation that the peak in non‑stabilizerness shifts slightly relative to the point of maximal extrinsic curvature when using a more realistic basis set does not forbid that the system can be tuned to exhibit very large values of magic without reaching the `risky' regime where the covalent bond irreversibly breaks, ensuring that the dimer remains intact and reusable. Such a controlled extraction and subsequent utilization of non‑stabilizerness could open up novel avenues for quantum simulation and computation.

    More concretely, in the minimal-basis H$_2$ setting studied here, the two dominant determinants define an encoded qubit, and the high-magic point near $\theta=-\pi/8$ corresponds to a state that is Clifford-equivalent to a T-type magic resource. In that encoded picture, a coherently prepared stretched dimer could therefore be viewed, at least in principle, as a non-stabilizer ancilla enabling a standard state-injection primitive for the implementation of a T gate on a computational mode using Clifford operations, measurements, and Pauli feedforward. By contrast, an uncontrolled relaxation would simply dissipate the stored non-stabilizerness and should not be interpreted as a computational primitive.

\section*{Acknowledgements}

    M.S. would like to thank Junggi Yoon and Pablo Martinez Azcona for discussions, and Kyung Hee University for visits during the writing of this manuscript. The authors acknowledge funding via the FNR-CORE Grant ``BroadApp'' (FNR-CORE C20/MS/14769845) and ERC-AdG Grant ``FITMOL''. We thank the anonymous referees for their thorough reading of the manuscript and for their insightful comments, which allowed us to enhance both the clarity and the overall presentation of this work.

\section{Code and Data Availability}The reader will find an open source python code to reproduce the results of this paper at the following public GitHub repository \cite{MSrepo}.

\bibliographystyle{unsrt}
\bibliography{main}

\end{document}